# A TRANSVERSE CURRENT RECTIFICATION IN GRAPHENE SUPERLATTICE

*D. V. Zav'yalov, V. I. Konchenkov, S. V. Kryuchkov*

Volgograd State Pedagogical University

400131, Russia, Volgograd, Lenina av., 27

sed@fizmat.vspu.ru

A model for energy spectrum of superlattice on the base of graphene placed on the striped dielectric substrate is proposed. A direct current component which appears in that structure perpendicularly to pulling electric field under the influence of elliptically polarized electromagnetic wave was derived. A transverse current density dependence on pulling field magnitude and on magnitude of component of elliptically polarized wave directed along the axis of a superlattice is analyzed.

## 1. Introduction

Graphene was experimentally obtained in 2004 [1] and now it attracts an attention of scientists because of ability to observe new effects which are typical of only this material (see, for example, [2, 3, 4]) and because of probable applications in microelectronics (see, for example, [5, 6]). In recent time an attention of researchers is concentrated also on studying of artificial structures which may be created on the base of graphene and particularly on studying of superlattices on the base of graphene (see [7] and references in this paper). An important feature of graphene energy spectrum [8] is its nonadditivity. A formation energy zones with finite widths is possible if additional periodical potential was shaped and it may lead to occurrence of some nonlinear effects which are typical to narrow-band semiconductors. The phenomena associated with interdependence of motions of charge carriers along mutually transverse directions are possible in concerned structure if nonadditivity of spectrum is maintained. So superlattices on the base of graphene are of interest for research.

## 2. Approximated expression of an energy spectrum of a superlattice on the base of graphene.

In [7] a model of superlattice on the base of graphene on spripped dielectric substrate is proposed. Ribbons of so-named gap and gapless modification of this material are the layers of a superlattice. An energy gap appears because of interaction between graphene layer and substrate on which the sample is placed. If one makes a substrate consisted of alternate layers of different dielectrics and puts graphene on it any way, a periodical potential along the direction of interchange of layers appears. This structure is shown in figure 1 schematically.

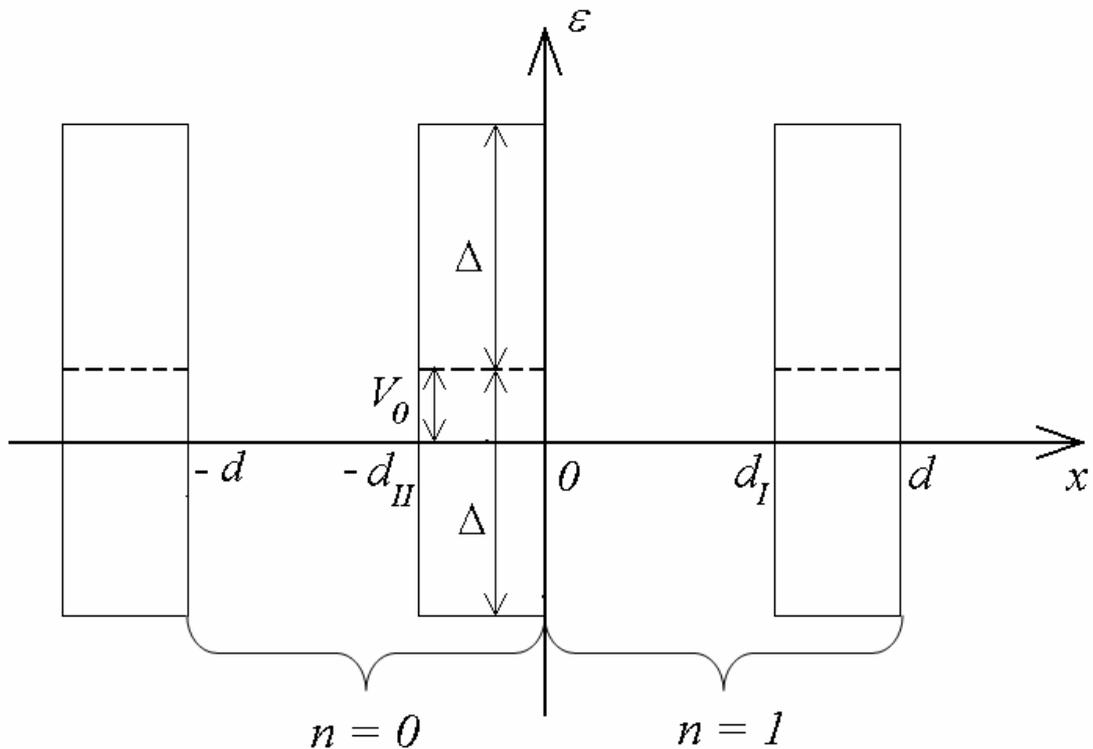

*Figure 1. A model of superlattice consisted of interlaced strips of gap and gapless graphene (it presented in [7])*

Here $\Delta$ is a halfwidth of a forbidden zone of gap modification of graphene, $V_0$ is an additional scalar potential which may arise because of mismatch of a Dirac point position of gapless modification of graphene and a midpoint of a forbidden zone of gap modification, $d_I$ and $d_{II}$ are widths of gapless and gap modification of graphene respectively, $d = d_I + d_{II}$ is a superlattice period. In a paper [7] a dispersion relation for considering structure was derived:

$$\begin{cases} \dfrac{(B^2 q_y^2 - E^2 + EV)}{\sqrt{a_1}\sqrt{b_1}} \operatorname{sh}\left(\dfrac{b\sqrt{b_1}}{B}\right)\sin\left(\dfrac{a\sqrt{a_1}}{B}\right) + \operatorname{ch}\left(\dfrac{b\sqrt{b_1}}{B}\right)\cos\left(\dfrac{a\sqrt{a_1}}{B}\right) - \cos q_x = 0, \ b_1 \geq 0; \\ \dfrac{(B^2 q_y^2 - E^2 + EV)}{\sqrt{a_1}\sqrt{-b_1}} \sin\left(\dfrac{b\sqrt{-b_1}}{B}\right)\sin\left(\dfrac{a\sqrt{a_1}}{B}\right) + \cos\left(\dfrac{b\sqrt{-b_1}}{B}\right)\cos\left(\dfrac{a\sqrt{a_1}}{B}\right) - \cos q_x = 0, \ b_1 < 0. \end{cases} \quad (1)$$

Here $a_1 = E^2 - B^2 q_y^2$, $b_1 = 1 + B^2 q_y^2 - (E-V)^2$, $B = (\hbar v_f)/(\Delta d)$, $E = \varepsilon/\Delta$, $V = V_0/\Delta$, $q_y = k_y d$, $q_x = k_x d$, $a = d_1/d$, $b = d_2/d$, $\varepsilon$ is an energy of electron, $k_x$, $k_y$ are the components of quasiwave vector of electron, $v_f$ is a Fermi velocity in graphene. Note that similar dispersion relation was obtained in [9] so.

The dispersion relation (1) was numerically solved for $E$ and then approximate formula for functions $E_n(q_x, q_y)$ (n is a number of minizone) are fitted for farther analytical investigations in this work. The Levenberg-Marquardt algorithm (LMA) (see, for example, [10]), which is realized in mathematical package Wolfram Mathematica, was used for finding the coefficients. In wide ranges of values of parameters $a$, $b$, $B$ and $V$ the next expression is a good approximation for electron energy in minizone:

$$E_0 = f_1 + \sqrt{f_2^2 + f_3^2 q_y^2} + \frac{f_4(1 - \cos q_x)}{\sqrt{f_2^2 + f_3^2 q_y^2}}. \quad (2)$$

Here coefficients $f_i$, $i = 1,2,3,4$ are fitted numerically. A graphic of relative error $\eta = \left|\dfrac{E - E_0}{E_0}\right|$ which one introduces because of replacement of the energy values $E$ gotten directly from the numerical solution of dispersion relation by the energy values $E_0$ derived from formulae (2) is

shown in figure 2. The relative error is not more then 2 per cent. Coefficients $f_i$ using for calculations correspond to one-minizone approximation. Under the condition $V_0 = 0$ this approximation is correct at the next parameters of superstructure:

$$a = b = 0.5, \quad B = 0.25. \tag{3}$$

In this case the coefficients $f_i$ take next values:

$$f_1 = -0.0273612, \quad f_2 = 0.451551, \quad f_3 = 0.255208, \quad f_4 = 0.0483906. \tag{4}$$

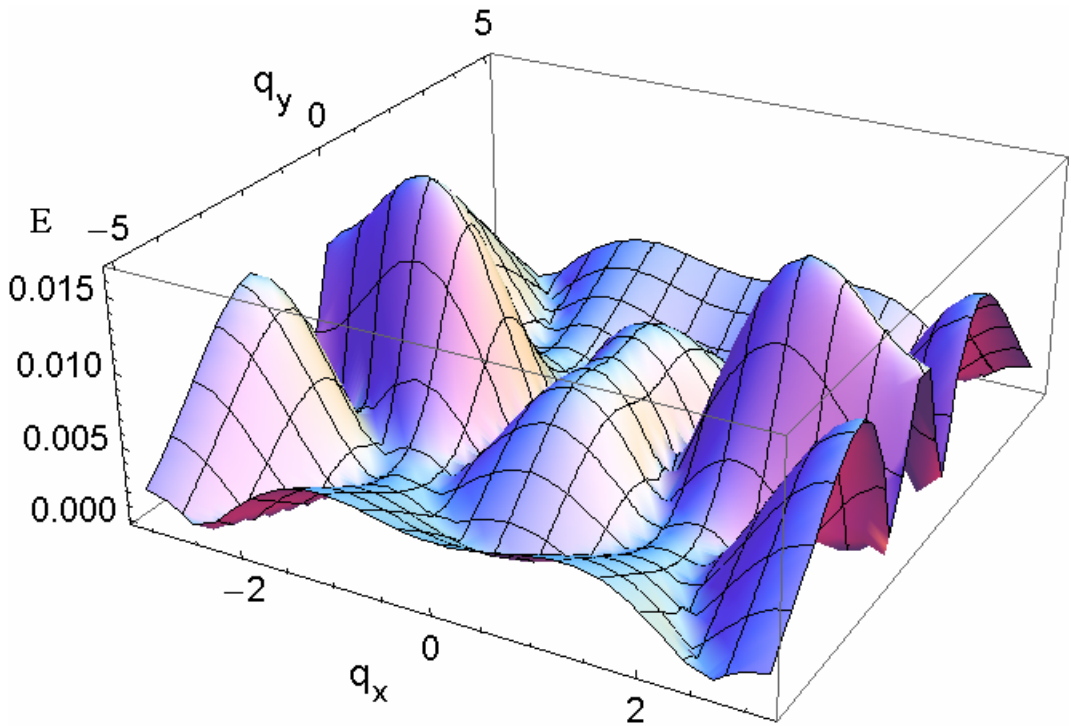

*Figure 2. Graphic of relative difference between energy value which was gotten from numerical solution of dispersion relation and energy values calculated with model energy spectrum.*

Two first minizones where electron energy is more than energy in Dirac point of gapless modification of graphene are presented in figure 3.

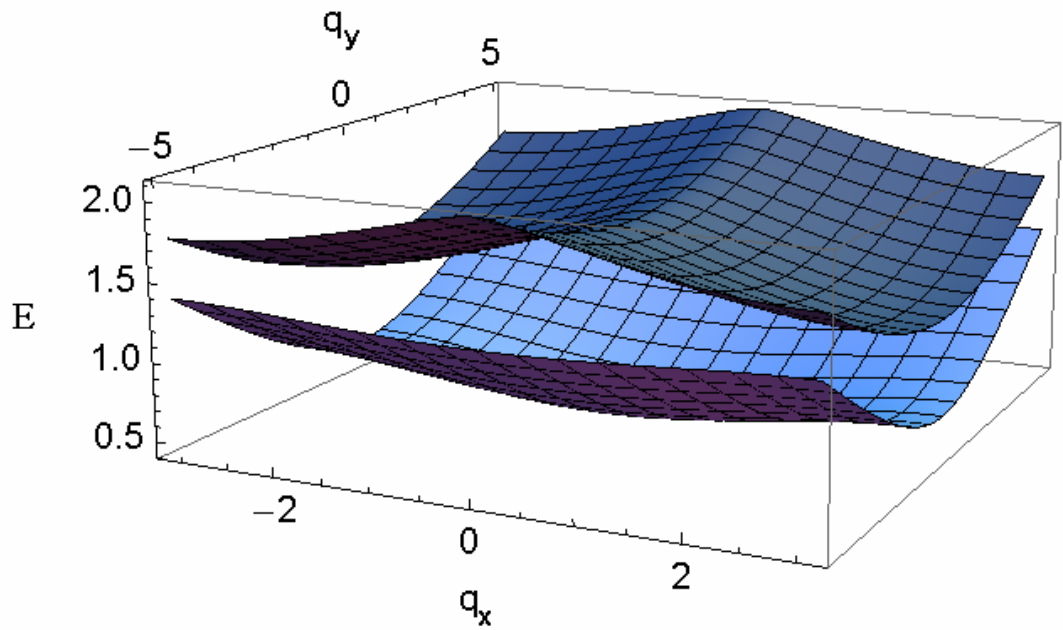

**Figure 3.** *Energy dependencies on the quasiwave vector in two first energy minizones.*

Graphics of dependence $E_n(q_x)$ for four nearest to Dirac point minizones under the condition $q_y = 0$ are shown in figure 4.

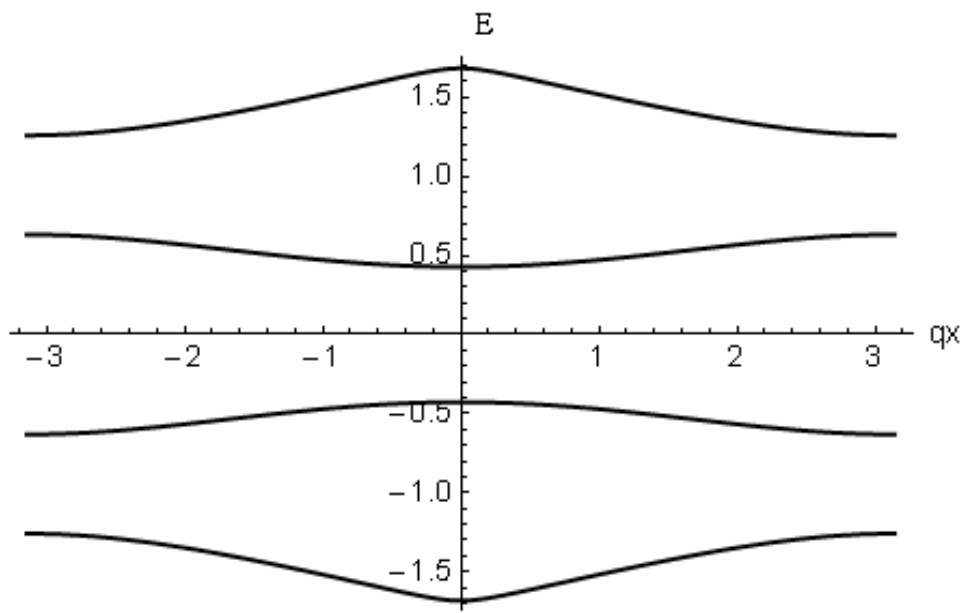

**Figure 4.** *Energy dependencies on the quasiwave vector component $q_x$ at $q_y = 0$.*

A width of forbidden zone between the first and the second minizones $\varepsilon_{1,2}^g$ is approximately equal to $0.6\Delta$, a width of the first minizone $\Delta_1$ is equal to $\Delta_1 \approx 0.2\Delta$, an energy gap between lowermost electron zone and topmost hole zone $\varepsilon_0^g$ is equal to $\varepsilon_0^g \approx 0.85\Delta$.

In the paper [11] another expression for energy spectrum of superlattice on the base of graphene is proposed. In our denotation this expression takes the next form:

$$E = \sqrt{1 + B^2 q_y^2 + 2B^2(1 - \cos q_x)}. \qquad (5)$$

The expression for energy spectrum (2) is more convenient for analytical investigations so in present paper it is used.

In the capacity of gapless modification of graphene we shall propose graphene on the substrate of $SiO_2$ [1]. Several modification of graphene have a forbidden zone in their energy spectrum. They are epitaxial graphene on the silicon carbide substrate $SiC$ [12] the bandgap width of which is equal to $2\Delta \approx 0.26\,eV$ and the sample placed over the hexagonal lattice of boron nitride $h-BN$ [13] with forbidden zone width $2\Delta \approx 0.053\,eV$. To additional periodical potential impacts on motion of particle the period of superstructure $d$ should be much less than mean free path $\lambda$: $d \ll \lambda$ [14]. The mean free path in graphene at room temperature and at charge carrier concentration $n \sim 10^{12}\,cm^{-2}$ is about $\lambda \sim 4 \cdot 10^{-5}\,cm$ [1]. For equality $B = 0.25$ in superlattice with graphene over boron nitride substrate in the capacity of gap modification the period of superstructure $d$ should be equal approximately 650 interatomic distance ($d \sim 0.2\lambda$) but using graphene on silicon carbide wafer the period of superstructure should be about 135 interatomic distances ($d \sim 0.05\lambda$). In view of that one-minizone approximation may be use only in superlattice on the base of graphene placed on the substrate consisted of interlaced strips of silicon oxide and silicon carbide.

### 3. Formulation of problem.

The energy spectrum derived by expression (2) as well as spectrum of graphene has a feature to be non-additive. Nonadditivity of energy spectrum as it was noted may lead to appearance of so-named current rectification effects. In a paper [11] a problem of mutual rectification of currents induced by two sinusoidal waves with transverse each other planes of polarization in superlattice on the base of graphene with energy spectrum in form (5) was investigated. In [11] it was shown that direct current component appears along the direction of electric field strength vector of wave with a frequency twice as much the frequency of another wave and that the current density dependence on magnitudes of incident waves has nonlinear character. Previously in a paper [15] a direct component of current density was derived in graphene in a direction perpendicular to pulling electric field in situation when elliptically polarized electromagnetic wave is incident normally to the surface of the sample. In [16] the effect studied in [15] was modeled using Monte Carlo simulation. An appearance of direct current in graphene under influence of elliptically polarized laser radiation was experimentally observed in [17, 18].

Ones of peculiarities of spectrum (2) is its periodicity on $q_x$ and a dependence of allowed band width of superlattice which was built along X axis on a value of $q_y$. It can be expected on the one hand, the nonlinear dependence of current density on electric field which is proper to the superlattice and on the other hand, in connection with the nonadditivity of the spectrum the appearance of constant current in the direction along which there is no static electric field. So it is interesting to study a problem which is similar to the problem investigated in [15] in superlattice on the base of graphene for examinination of influence of periodical potential on electronic properties of graphene.

A problem geometry is shown in figure 4.

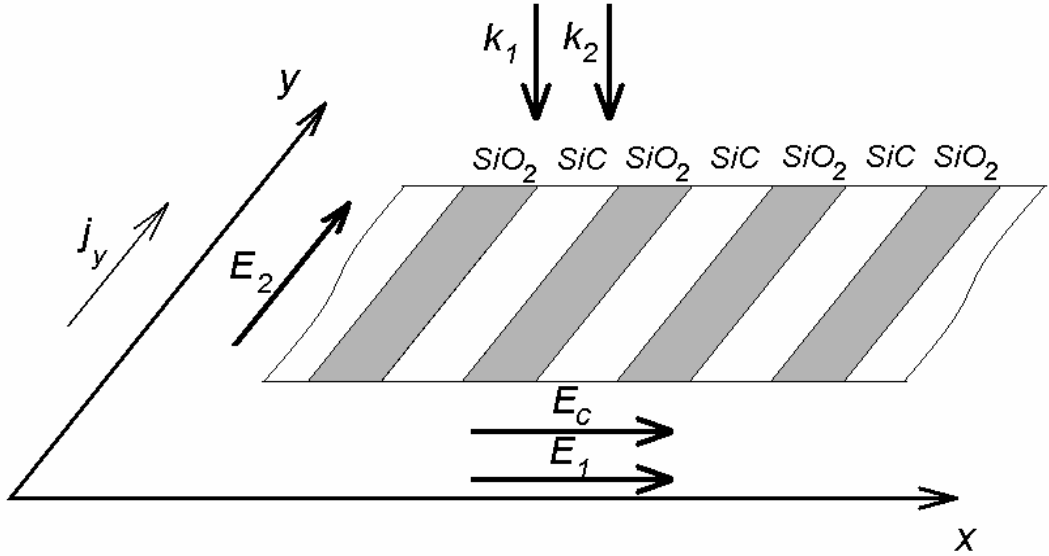

*Figure 5. Problem geometry.*

Here $E_c$ is a constant field strength, $E_1 = E_{10}\cos(\omega t)$, $E_2 = E_{20}\cos(\omega t + \varphi)$ are components of electric field strength of elliptically polarized wave. We shall find a direct component of current density along $Oy$ axis in approach of constant relaxation frequency $\nu$:

$$j_y = \left\langle e\nu \sum_{\mathbf{p}} \int_{-\infty}^{t} dt_1 \exp(-\nu(t-t_1)) v_y\left(\mathbf{p} + \frac{e}{c}\mathbf{A}\right) f_0(\mathbf{p}) \right\rangle_t. \tag{6}$$

Here $A = -c\int E\,dt$ is a vector potential of result electric field, $f_0(\mathbf{p})$ is the equilibrium Boltzmann distribution function, $v_y(\mathbf{p}) = \partial\varepsilon/\partial p_y$ is a velocity component, $e$ is a value of elementary charge, angle brackets $\langle\ \rangle_t$ denote an averaging by large in comparing with a period of incident wave time interval. We introduce next designations: $F_0 = \dfrac{eE_c d}{\hbar\omega}$, $F_1 = \dfrac{eE_{10}d}{\hbar\omega}$, $F_2 = \dfrac{eE_{20}d}{\hbar\omega}$, $T_1 = \dfrac{T}{\Delta}$, $\gamma = \dfrac{\nu}{\omega}$. The expression of current density takes a form:

$$j_y = \frac{j_0}{2\pi f_4 P_0}\left\langle \int_{-\infty}^{0} dt_1 \exp(\gamma t_1)\int_{-\infty}^{\infty} dq_y \int_{-\pi}^{\pi} dq_x \frac{q_y - F_2(\sin(t_1 + t + \varphi) - \sin(t + \varphi))}{\sqrt{f_2^2 + f_3^2(q_y - F_2(\sin(t_1 + t + \varphi) - \sin(t + \varphi)))^2}}\right.$$

$$\left.\cdot\left(1 - \frac{f_4(1 - \cos(q_x - F_0 t_1 - F_1(\sin(t_1 + t) - \sin t)))}{f_2^2 + f_3^2(q_y - F_2(\sin(t_1 + t + \varphi) - \sin(t + \varphi)))^2}\right)\exp\left(-\frac{f_2^2 + f_3^3 q_y^2 + f_4(1 - \cos q_x)}{T_1\sqrt{f_2^2 + f_3^3 q_y^2}}\right)\right\rangle_t$$

(7)

Here $j_0 = en\dfrac{\Delta d}{\hbar}\gamma f_3^2 f_4$, $n$ is a surface concentration of charge carriers,

$$P_0 = \int_{-\infty}^{\infty} dq_y I_0\left(\dfrac{f_4}{T_1\sqrt{f_2^2 + f_3^2 q_y^2}}\right)\exp\left(-\dfrac{f_2^2 + f_3^3 q_y^2 + f_4}{T_1\sqrt{f_2^2 + f_3^2 q_y^2}}\right),$$ $I_0(x)$ is a modified Bessel function of first kind. In a first non-vanishing approximation by dimensionless field strength $F_2$ an expression of current density becomes

$$j_y = \dfrac{j_0}{P_0}F_2 \int_{-\infty}^{\infty} dq_y \dfrac{2f_3^2 q_y^2 - f_2^2}{(f_2^2 + f_3^2 q_y^2)^{5/2}} I_1\left(\dfrac{f_4}{T_1\sqrt{f_2^2 + f_3^2 q_y^2}}\right)\exp\left(-\dfrac{f_2^2 + f_3^3 q_y^2 + f_4}{T_1\sqrt{f_2^2 + f_3^2 q_y^2}}\right) \cdot$$
$$\cdot \left\langle \int_{-\infty}^{0} dt_1 \exp(\gamma t_1)\cos(F_0 t_1 + F_1(\sin(t_1+t)-\sin t))(\sin(t_1+t+\varphi)-\sin(t+\varphi))\right\rangle_t. \quad (8)$$

Finally obtain:

$$j_y = j_0 \dfrac{P_1}{P_0}\cos\varphi\, F_0 F_2 \sum_{n=-\infty}^{\infty} J_n(F_1)(J_{n-1}(F_1)-J_{n+1}(F_1))\dfrac{F_0^2 - n^2 + \gamma^2}{((F_0-n)^2+\gamma^2)((F_0+n)^2+\gamma^2)}, \quad (9)$$

.

where $P_1 = \int_{-\infty}^{\infty} dq_y \dfrac{2f_3^2 q_y^2 - f_2^2}{(f_2^2 + f_3^2 q_y^2)^{5/2}} I_1\left(\dfrac{f_4}{T_1\sqrt{f_2^2 + f_3^2 q_y^2}}\right)\exp\left(-\dfrac{f_2^2 + f_3^3 q_y^2 + f_4}{T_1\sqrt{f_2^2 + f_3^2 q_y^2}}\right).$

Graphics of dependence of direct current density on dimensionless field strengths $F_0$, $F_1$ are shown in figures 5, 6.

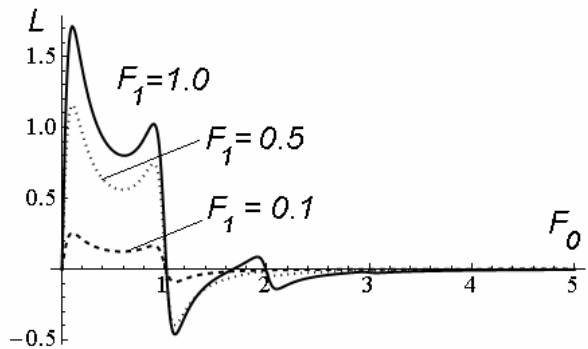
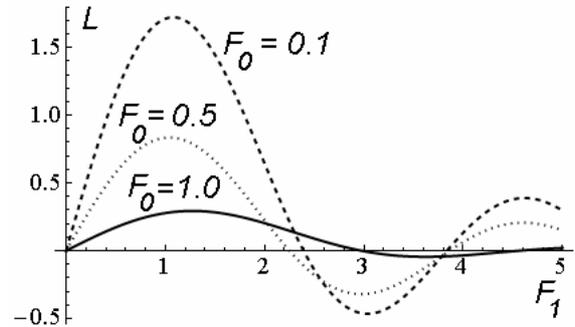

Figure 5.     Figure 6.

Note that these dependences have parts of negative differential conductivity.

At a temperature $T \sim 70\,K$, $\gamma \sim 0.1$, $\omega \sim 10^{12}\,c^{-1}$, $F_0, F_1 \sim 1\,SGSE\ units$, $F_2 \sim 0.1\,SGSE\ units$, $n \sim 10^{10}\,cm^{-2}$, $d \sim 10^{-6}\,cm$ a current density $j_y \sim 10^{-4}\,A/cm$.

**Acknowledgements.**

The work was supported by RFBR Project 10-02-97001-r_povolzhe_a and was carried out within the framework of the program "Development of the Scientific Potential of the Higher School".

**References.**

[1] K. S. Novoselov, A. K. Geim, S. V. Morozov, D. Jiang, Y. Zhang, S. V. Dubonos, I. V. Grigorieva, A. A. Firsov, "Electric Field Effect in Atomically Thin Carbon Films," *Science*, **306**, 666 (2004).

[2] A. H. Castro Neto, F. Guinea, N. M. R. Peres, K. S. Novoselov, A. K. Geim, "The electronic properties of graphene," *Rev. Mod. Phys.*, **81**, 109 (2009).

[3] Y. E. Lozovick, S. P. Merkulova and A. A. Sokolik, "Electronic transport in graphene", *Physics-Uspekhi,* **51** 727 (2008).

[4] S. V. Morozov, K. S. Novoselov and A. K. Geim "Collective electron phenomena in graphene", *Physics-Uspekhi,* **51** 744 (2008).

[5] Y. Q. Wu, P. D. Ye, M. A. Capano, Y. Xuan, Y. Sui, M. Qi, J. A. Cooper, T. Shen, D. Pandey, G. Prakash, R. Reifenberger, "Top-gated graphene field-effect-transistors formed by decomposition of SiC," *Applied Physics Letters*, **92**, 092192, (2008).

[6] Thomas Mueller, Fengnian Xia, Phaedon Avouris, "Graphene photodetectors for high-speed optical communications," *Nature Photonics*, **4**, 297 (2010).

[7] P. V. Ratnikov, "Superlattice based on graphene on a strip substrate," *JETP letters*, **90** (6). P 469 (2009).

[8] P. R. Wallace, "The Band Theory of Graphite," *Phys. Rev.* **71**, 622 (1947).

[9] Li-Gang Wang, Xi Chen, "Robust zero-averaged wave-number gap inside gapped graphene superlattices". *E-print: arXiv:1008.0504v1*.

[10] P. Gill, W. Murray, M. H. Wright, *Practical optimization*, (Academic Press Inc., London, 1981).

[11] S. V. Kryuchkov, E. I. Kuhar and V. A. Yakovenko, "Effect of mutual rectification rectification of two electromagnetic waves with perpendicular polarization planes in a superlattice based on graphene," *Bulletin of Russian Academy of Sciences: Physics*, **74** (12), P. 1679.

[12] S. Y. Zhou, G.-H. Gweon, A. V. Fedorov, P. N. First, W. A. de Heer, D.-H. Lee, F. Guinea, A. H. Castro Neto, A. Lanzara, "Substrate-induced band gap opening in epitaxial graphene," *Nature Materials*, **6**, 770, (2007).

[13] Gianluca Giovannetti, Petr A. Khomyakov, Geert Brocks, Paul J. Kelly, and Jeroen van den Brink "Substrate-induced band gap in graphene on hexagonal boron nitride: Ab initio density functional calculations," *Phys. Rev. B*, **76**, 073103 (2007).

[14] F. G. Bass, A. A. Bulgakov, A. P. Tetervov, *High-frequency properties of semiconductors with superlattices* (Nauka, Moskow, 1989).

[15] D. V. Zav'yalov, S. V. Kryuchkov, É. V. Marchuk, "On the possibility of transverse current rectification in graphene," *Technical Physics Letters*, **34**(11), 915 (2008).

[16] D. V. Zav'yalov, S. V. Kryuchkov, T. A. Tyulkina, "Effect of rectification of current induced by an electromagnetic wave in graphene: A numerical simulation," *Semiconductors*, **44**(7), 879 (2010).

[17] J. Karch, P. Olbrich, M. Schmalzbauer, C. Brinsteiner, U. Wurstbauer, M. M. Glazov, S. A. Tarasenko, E. L. Ivchenko, D. Weiss, J. Eroms, S. D. Ganichev, "Photon helicity driven electric currents in graphene," *E-print: arXiv:1002.1047v1 (2010)*.

[18] J. Karch, P. Olbrich, M. Schmalzbauer, C. Zoth, C. Brinsteiner, M. Fehrenbacher, U. Wurstbauer, M. M. Glazov, S. A. Tarasenko, E. L. Ivchenko, D. Weiss, J. Eroms , R. Yakimova, S. Lara-Avila, S. Kubatkin, S. D. Ganichev, "Circular ac Hall Effect," *E-print: arXiv:1008.2116v1 (2010)*.